\pdfoutput=1
\documentclass[sigconf]{acmart}
\usepackage{color}
\usepackage{bbm}
\usepackage{multirow}
\usepackage[inline]{enumitem}
\usepackage{subfigure} 
\usepackage{sistyle}
\usepackage[ruled]{algorithm2e}
\usepackage{booktabs}
\usepackage{caption}
\SIthousandsep{,}
\usepackage{makecell}
\usepackage[skip=0pt]{caption}
\usepackage{amsmath}
\usepackage{hyperref}
\usepackage{array} 
\usepackage{graphicx}
\usepackage{xcolor}
\usepackage{acronym}
\usepackage[export]{adjustbox}

\newcommand{\heading}[1]{\vspace*{0.5mm}\noindent\textbf{#1:}}

\AtBeginDocument{%
  \providecommand\BibTeX{{%
    \normalfont B\kern-0.5em{\scshape i\kern-0.25em b}\kern-0.8em\TeX}}}

\makeatletter
\g@addto@macro\normalsize{%
  \abovedisplayskip 2pt plus1pt 
  \belowdisplayskip 2pt plus1pt
  \abovedisplayshortskip  2pt plus1pt%
  \belowdisplayshortskip  1pt plus1pt
}
\makeatother

\acrodef{IR}{information retrieval}
\acrodef{LLM}{large language model}

\AtBeginDocument{%
  \providecommand\BibTeX{{%
    Bib\TeX}}}
\setcopyright{acmlicensed}
\copyrightyear{2018}
\acmYear{2018}
\acmDOI{XXXXXXX.XXXXXXX}
\acmConference[Conference acronym 'XX]{Make sure to enter the correct
  conference title from your rights confirmation email}{June 03--05,
  2018}{Woodstock, NY}
\acmISBN{978-1-4503-XXXX-X/18/06}

\author{Hengran Zhang}
\affiliation{
\institution{State Key Laboratory of AI Safety, ICT, CAS}
  \institution{University of Chinese Academy of Sciences}
  \city{Beijing}
  \country{China}
}
\email{zhanghengran22z@ict.ac.cn}

\author{Minghao Tang}
\affiliation{
\institution{State Key Laboratory of AI Safety, ICT, CAS}
  \institution{University of Chinese Academy of Sciences}
  \city{Beijing}
  \country{China}
}
\email{tangminghao25s@ict.ac.cn}

\author{Keping Bi}
\affiliation{
\institution{State Key Laboratory of AI Safety, ICT, CAS}
  \institution{University of Chinese Academy of Sciences}
  \city{Beijing}
  \country{China}
}
\email{bikeping@ict.ac.cn}

\author{Jiafeng Guo}
\affiliation{
\institution{State Key Laboratory of AI Safety, ICT, CAS}
  \institution{University of Chinese Academy of Sciences}
  \city{Beijing}
  \country{China}
}
\email{guojiafeng@ict.ac.cn}

\renewcommand{\shortauthors}{Hengran Zhang et al.}
\begin{document}

\title{Beyond Relevance: Utility-Centric Retrieval in the LLM Era}

\begin{abstract}

Information retrieval systems have traditionally optimized for topical relevance—the degree to which retrieved documents match a query. However, relevance only approximates a deeper goal: utility, namely, whether retrieved information helps accomplish a user's underlying task. The emergence of retrieval-augmented generation (RAG) fundamentally changes this paradigm. Retrieved documents are no longer consumed directly by users but instead serve as evidence for large language models (LLMs) that produce answers. As a result, retrieval effectiveness must be evaluated by its contribution to generation quality rather than by relevance-based ranking metrics alone. 
This tutorial argues that retrieval objectives are evolving from relevance-centric optimization toward LLM-centric utility. We present a unified framework covering LLM-agnostic versus LLM-specific utility, context-independent versus context-dependent utility, and the connection with LLM information needs and agentic RAG. By synthesizing recent advances, the tutorial provides conceptual foundations and practical guidance for designing retrieval systems aligned with the requirements of LLM-based information access.
\end{abstract}


\maketitle
\acresetall
\section*{Cover Sheet Information}
\heading{Title} Beyond Relevance: Utility-Centric Retrieval in the LLM Era


\heading{Format} This tutorial is a half-day (3-hour) lecture-style session that includes scheduled breaks and will be conducted on-site.

\heading{Intended Audience} 
Intermediate. This tutorial targets an audience who is interested in RAG and LLM-based information access. It is particularly relevant to IR researchers exploring utility-oriented retrieval and to NLP/LLM researchers seeking principled ways to integrate retrieval into generative systems. 

\heading{Prerequisite Knowledge} Basic familiarity with information retrieval and large language models is assumed.

\heading{Previous Talks} This tutorial has not been presented before.


\heading{Presentors}

\textbf{Hengran Zhang} is a Ph.D. student at the Institute of Computing Technology (ICT), Chinese Academy of Sciences (CAS). Her research centers on utility-focused RAG and dense retrieval. Her work has been published in top-tier conferences, including SIGIR and EMNLP. She has conducted extensive research on 
utility-focused RAG, including utility-based evidence selection \cite{zhang2024large, zhang2025distilling, zhang2024iterative, zhang2023relevance}, utility-focused annotation for retrieval and RAG \cite{zhang2025leveraging}, and LLM-specific utility modeling \cite{zhang2025llm}.

\textbf{Minghao Tang} is a master’s student at the ICT, CAS. His research focuses on knowledge injection for LLMs. He has collaborated with Hengran Zhang and Keping Bi on utility-focused RAG \cite{zhang2025leveraging} and has publications at EMNLP, ICLR, and SIGIR-AP. 

\textbf{Keping Bi} is an Associate Professor at the ICT, CAS. Her research primarily focuses on trustworthy information access, with special emphasis on IR and retrieval-augmented generation. She received her Ph.D. from the University of Massachusetts Amherst under the supervision of Prof. W. Bruce Croft. Dr. Bi has served in several academic leadership roles, including General Co-Chair of SIGIR-AP 2025 and Co-Editor of SIGIR Forum (2025-2028). 

\textbf{Jiafeng Guo} is a Professor at the ICT, CAS. His current research focuses on agentic information seeking, generative retrieval, and retrieval robustness. He has received multiple Best Paper Awards at top-tier conferences, including CIKM, SIGIR, and EMNLP. He has also actively contributed to the IR community through conference organization, serving as Short Paper Co-Chair of SIGIR 2020 and ICTIR 2019, and PC Co-Chair of CCIR 2018, among other roles.
He has delivered tutorials on generative retrieval and robust IR at major conferences, including SIGIR, WSDM, and CIKM.

\section{Motivation and Overview}

Relevance and utility have long been recognized as two foundational criteria in information retrieval \cite{saracevic1975relevance, saracevic1996relevance, saracevic1988study}. Relevance typically refers to topical relevance, the degree of correspondence between the subject of a query and retrieved documents, often operationalized through the notion of ``aboutness''  \cite{saracevic1988study}. In contrast, utility concerns the usefulness or value of retrieved information to an information seeker, emphasizing whether retrieved items help achieve the user’s underlying goals \cite{cooper1971definition, saracevic1975relevance, saracevic1988study}.
Because usefulness depends on user context, task, and situational needs, utility has historically been difficult to observe and quantify automatically \cite{bruce1994cognitive}. Topical relevance, by comparison, is more straightforward to annotate and evaluate, making it the dominant optimization target in IR research for decades, especially dense retrieval studies in recent years \cite{karpukhin2020dense, zhang2025unleashing}.

The rise of retrieval-augmented generation (RAG) introduces a fundamental shift. Retrieval results are no longer primarily presented to human users as ranked lists; instead, they serve as evidence consumed by large language models (LLMs) that synthesize answers. This transformation changes the role of retrieval systems. Effectiveness is no longer determined by whether documents appear relevant to human readers or help them achieve their goals, but by whether they improve the quality of generated responses \cite{zhang2024large}. Under this setting, a document may be topically relevant yet ineffective for LLM generation, while other evidence may substantially enhance reasoning, grounding, or factual correctness. Importantly, compared with user-centric utility, LLM-centric utility is often more directly measurable through answer-level metrics such as correctness or likelihood-based evaluation, making it a more tractable and feasible optimization objective.

This tutorial argues that the objective of information retrieval is evolving, from topical relevance and user-centric utility,  toward LLM-centric utility. By tracing this progression, we provide a unified perspective on how retrieval objectives are changing in the LLM era and review emerging work on modeling, estimating, and optimizing LLM-centric utility.

\subsection{User-Centric Utility}

As early as 1971, \citet{cooper1971definition} distinguished topical relevance from utility, arguing that relevance assesses whether information pertains to the subject of a query from a system-oriented perspective, whereas utility reflects the ultimate usefulness of information to the user. The notion of utility was later considered as a type of relevance in the debates about what relevance truly represents in information retrieval \cite{saracevic1996relevance, saracevic1975relevance, schutz2011reflections}. This user-oriented perspective for retrieval emphasizes that usefulness depends on a user’s situation, intent, and task context. Information becomes valuable not merely because it matches a topic, but because it contributes to problem resolution or decision-making.

Because utility cannot be directly observed, practical systems developed indirect ways to approximate it. Web search engines increasingly relied on implicit feedback signals, such as clicks, dwell time, and session behavior, to infer usefulness at scale \cite{jung2007click, kelly2004display, buscher2009segment, zhu2012more, lu2018between, luo2017does}. In recommender systems, measures such as clicks and purchases are often used as optimization objectives, as they more directly reflect user satisfaction and task success \cite{yu2016dynamic}. Furthermore, modern ranking systems frequently optimize the utility of the entire ranked list rather than treating documents independently at each position \cite{wang2016beyond,dai2020u}. This perspective moves beyond the traditional probabilistic ranking principle \cite{robertson1977probability}, which focuses on ranking documents solely according to individual relevance probabilities, and instead emphasizes holistic outcome-driven optimization.


In this paradigm, humans simultaneously act as both consumers and evaluators of retrieval results. Retrieval systems learn from user behavior and are ultimately judged by how well they satisfy human information needs. User-centric utility thus represents a transition from static relevance estimation to interaction-aware and outcome-oriented information access.

\vspace{-1mm}
\subsection{LLM-Centric Utility}

RAG fundamentally reshapes the role of retrieval in information access by introducing a new consumer of retrieval results: the LLM itself. Instead of directly assisting human readers, retrieved documents now function as supporting evidence that guides generation. The quality of retrieval must therefore be assessed through its contribution to generation outcomes rather than through standalone document relevance.

Traditional retrieval metrics, such as nDCG, MAP, and MRR, measure ranking quality with respect to topical relevance. However, they do not directly capture whether retrieved evidence improves generated answers \cite{canale-etal-2025-bes4rag}. In contrast to web search, where users directly judge retrieved results, RAG systems decouple evaluation and consumption: users assess the final synthesized answer, while retrieval utility is defined by how effectively evidence supports the LLM’s generation process. 
In this tutorial, we define LLM-centric utility as the extent to which retrieved information improves generation quality, such as answer correctness. We organize related work into three complementary dimensions:

\textbf{LLM-Agnostic Utility and LLM-Specific Utility}.
LLM-agnostic utility assumes that evidence possesses intrinsic informational value that generalizes across different generators. Unlike human-annotated topical relevance, utility has been approximated by measuring generation performance conditioned on specific documents, for example, using BLEU, ROUGE, EM, or F1 against ground-truth answers \cite{hofstatter2022multi, gan2024similarity, gao2024preference, ke2024bridging}; the likelihood of the ground-truth answer given the query and document \cite{shi2023replug, lewis2020retrieval}; model attention distributions over input documents \cite{izacard2022few, izacard2020distilling}; or performance differences between using and not using a document \cite{zhang2023relevance, dai2025seper, hu2023read, zhao2024seer, zhang2023relevance}. 
To obtain utility signals without relying on ground-truth answers, utility can also be annotated by advanced LLMs based on pseudo answers they generate, enabling scalable estimation of document usefulness without human labeling \cite{zhang2024iterative, zhang2024large, zhang2025distilling, zhang2025leveraging}.

In contrast, LLM-specific utility \cite{zhang2025llm} recognizes that the same evidence may vary in usefulness across different LLMs. Analogous to personalized web search, where identical results differ in value for users with distinct intentions, expertise levels, or constraints, LLMs may adopt different utility criteria due to variations in training data, internal knowledge, reasoning strategies, and comprehension capabilities \cite{zhang2025llm}.
In contrast to general utility that typically offers stronger cross-model generalization but may not maximize performance for a particular LLM, LLM-specific utility can be tailored to optimize generation quality for a given model, potentially improving performance at the cost of reduced transferability.

\textbf{Context-Independent and Context-Dependent Utility}.
Most existing approaches assume context-independent utility, where the usefulness of each document is evaluated independently of others \cite{shi2023replug, zhang2024large}. This simplification makes labeling and optimization tractable, allowing standard supervised learning techniques to be applied.
However, in realistic RAG settings, utility is often context-dependent. The usefulness of a document may depend on which other documents are present \cite{jain2025modeling}. A document may provide redundant information in one context but supply a crucial missing reasoning step in another. This notion resembles search result diversification, where ranking decisions consider both relevance and novelty.
When modeling context-dependent utility, the objective becomes retrieving a set of evidence that jointly maximizes overall generation utility. This requires capturing setwise interactions among documents and is particularly important in scenarios involving multi-aspect questions or multi-hop reasoning, where different pieces of evidence resolve different subproblems \cite{yang2018hotpotqa}.

\textbf{LLM Information Needs and Agentic RAG}. 
Beyond constructing utility labels and training retrievers accordingly, another line of work seeks to identify useful documents by explicitly modeling the LLM’s information needs. Rather than estimating document utility after retrieval, this perspective reframes retrieval as the process of satisfying the LLM’s latent knowledge gaps. Unlike web search, where users can iteratively reformulate queries as their information needs evolve, LLMs do not inherently express such needs. Consequently, researchers have proposed methods to infer and externalize these needs, for example, by generating queries based on model uncertainty reflected in low-confidence tokens or attention patterns over context \cite{jiang2023active, su2403dragin, jeong2024adaptive, shao2023enhancing, li2023llatrieval}.

A related paradigm is agentic RAG, in which the LLM iteratively generates queries, retrieves evidence, produces answers, and receives feedback based on answer quality. Reinforcement learning is then used to optimize reasoning and query generation policies \cite{jin2025search, gao2024smartrag, zheng2025deepresearcher, zhang2025evolvesearch}, with utility naturally emerging as the reward signal. However, in most agentic RAG systems, the retriever itself remains fixed, often relying on efficient but static methods such as BM25. While these approaches refine query generation and reasoning strategies, they typically do not reconsider the underlying retrieval objective. This tutorial, therefore, discusses agentic RAG in relation to utility-focused retrieval, while maintaining its primary emphasis on how retrieval objectives themselves evolve in the LLM era.

\subsection{Summary}
In summary, the LLM era challenges long-standing assumptions about what retrieval systems should optimize. As retrieval transitions from serving human readers directly to supporting generative models, topical relevance becomes an insufficient objective. A broader framework centered on utility, particularly LLM-centric utility, offers a principled way to rethink retrieval evaluation, supervision, and optimization.
By clarifying distinctions between relevance, user-centric utility, and LLM-centric utility, and by organizing emerging work along axes of model specificity, contextual dependency, and information need modeling, this tutorial aims to provide both conceptual grounding and a roadmap for future research on utility-oriented retrieval in the LLM era.

\section{Objectives}
This tutorial aims to equip attendees with a clear understanding of how retrieval objectives are evolving in the LLM era. Specifically, participants will:
\begin{itemize}[leftmargin=*,itemsep=0pt,topsep=0pt,parsep=0pt]
\item Understand the shift in retrieval goals, distinguishing topical relevance, user-centric utility, and LLM-centric utility, and why relevance-based optimization is insufficient for RAG systems.

\item Learn frameworks for modeling LLM-centric utility, including LLM-agnostic vs. LLM-specific utility and context-independent vs. context-dependent utility.

\item Examine evaluation and training strategies that align retrieval with downstream generation quality rather than traditional ranking metrics alone.

\item Identify open challenges and research directions in utility-oriented retrieval and its integration with agentic RAG.
\end{itemize}
By the end of the tutorial, attendees will gain both conceptual clarity and practical guidance for designing next-generation retrieval systems aligned with LLM-based generation.

\section{Relevance to the IR Community} 


Information retrieval is undergoing a fundamental transition. While traditional IR research has centered on estimating and optimizing topical relevance, the rapid adoption of retrieval-augmented generation (RAG) systems is reshaping the role of retrieval. Retrieved documents are increasingly consumed not by human users directly, but by large language models (LLMs) that generate responses. This shift raises foundational questions about what retrieval systems should optimize and how effectiveness should be evaluated.

This tutorial is timely and relevant to the IR community as it connects classical theories of relevance and utility with emerging LLM-based paradigms. By synthesizing recent advances in utility modeling, retriever optimization, and evaluation beyond traditional ranking metrics, the tutorial provides both conceptual grounding and research directions. It aims to support IR researchers and practitioners in designing retrieval systems better aligned with the evolving landscape of LLM-driven information access.
\vspace{-1mm}
\section{Format and Schedule}
This tutorial is designed as a half-day (3-hour) lecture-style session with scheduled breaks. It will be delivered on-site, with at least two of the presenters attending the conference in person.

\begin{itemize}[leftmargin=*,itemsep=0pt,topsep=0pt,parsep=0pt]

    \item Introduction and Foundations (20 min)
    \begin{itemize}
        \item Tutorial overview and motivation
        \item Relevance vs. utility: conceptual foundations
        \item From user-centric to LLM-centric utility
    \end{itemize}
    
    \item LLM-Agnostic and LLM-Specific Utility (50 min)
    \begin{itemize}
        \item Modeling LLM-agnostic utility
        \item Defining LLM-specific utility
        \item Trade-offs, evaluation, and open challenges
    \end{itemize}

    \item Q\&A (10 min)

    \item Break (30 min)
    
    \item Context-Independent and Context-Dependent Utility (30 min)
    \begin{itemize}
        \item Independent utility estimation
        \item Setwise and interaction-aware utility modeling
        \item Optimization challenges and future directions
    \end{itemize}
    
    \item LLM Information Needs and Agentic RAG (30 min)
    \begin{itemize}
        \item Modeling LLM information needs
        \item Query generation and agentic retrieval loops
        \item Reinforcement learning and system-level optimization
    \end{itemize}
    
    \item Q\&A and Discussion (10 min)

\end{itemize}

\vspace{-1mm}
\section{Tutorial Material} 
All supplemental materials for this tutorial will be accessible via a dedicated website shared with attendees beforehand. Resources include a comprehensive manuscript, a topic-organized reference list, and all tutorial slides. Additionally, a GitHub repository will be maintained to survey and categorize relevant research papers, supporting ongoing learning and exploration. 

\bibliographystyle{ACM-Reference-Format}
\balance
\bibliography{reference}

@inproceedings{zhang2024large,
  title={Are Large Language Models Good at Utility Judgments?},
  author={Zhang, Hengran and Zhang, Ruqing and Guo, Jiafeng and de Rijke, Maarten and Fan, Yixing and Cheng, Xueqi},
  booktitle={SIGIR'24},
  pages={1941--1951},
  year={2024}
}

@inproceedings{shi2023replug,
    title = "{REPLUG}: Retrieval-Augmented Black-Box Language Models",
    author = "Shi, Weijia  and
      Min, Sewon  and
      Yasunaga, Michihiro  and
      Seo, Minjoon  and
      James, Richard  and
      Lewis, Mike  and
      Zettlemoyer, Luke  and
      Yih, Wen-tau",
    booktitle = "NAACL'24",
    month = jun,
    year = "2024",
    pages = "8371--8384",
    
}

@article{saracevic1975relevance,
  title={Relevance: A review of and a framework for the thinking on the notion in information science},
  author={Saracevic, Tefko},
  journal={JASIST},
  volume={26},
  number={6},
  pages={321--343},
  year={1975},
  publisher={Wiley Online Library}
}

@inproceedings{saracevic1996relevance,
  title={Relevance reconsidered},
  author={Saracevic, Tefko},
  booktitle={Proceedings of the second conference on conceptions of library and information science (CoLIS 2)},
  pages={201--218},
  year={1996}
}

@article{bruce1994cognitive,
  title={A cognitive view of the situational dynamism of user-centered relevance estimation},
  author={Bruce, Harry W},
  journal={JASIST},
  volume={45},
  number={3},
  pages={142--148},
  year={1994},
  publisher={Wiley Online Library}
}

@article{cooper1971definition,
  title={A definition of relevance for information retrieval},
  author={Cooper, William S},
  journal={Information storage and retrieval},
  volume={7},
  number={1},
  pages={19--37},
  year={1971},
  publisher={Elsevier}
}

@article{saracevic1988study,
  title={A study of information seeking and retrieving. I. Background and methodology},
  author={Saracevic, Tefko and Kantor, Paul and Chamis, Alice Y and Trivison, Donna},
  journal={JASIST},
  volume={39},
  number={3},
  pages={161--176},
  year={1988},
  publisher={Wiley Online Library}
}

@book{schutz2011reflections,
  title={Reflections on the Problem of Relevance},
  author={Schutz, Alfred and Embree, Lester},
  year={2011},
  publisher={Springer}
}

@inproceedings{kelly2004display,
  title={Display time as implicit feedback: understanding task effects},
  author={Kelly, Diane and Belkin, Nicholas J},
  booktitle={SIGIR'04},
  pages={377--384},
  year={2004}
}

@inproceedings{buscher2009segment,
  title={Segment-level display time as implicit feedback: a comparison to eye tracking},
  author={Buscher, Georg and Van Elst, Ludger and Dengel, Andreas},
  booktitle={SIGIR'09},
  pages={67--74},
  year={2009}
}

@article{jung2007click,
  title={Click data as implicit relevance feedback in web search},
  author={Jung, Seikyung and Herlocker, Jonathan L and Webster, Janet},
  journal={Information processing \& management},
  volume={43},
  number={3},
  pages={791--807},
  year={2007},
  publisher={Elsevier}
}

@inproceedings{karpukhin2020dense,
  title={Dense Passage Retrieval for Open-Domain Question Answering.},
  author={Karpukhin, Vladimir and Oguz, Barlas and Min, Sewon and Lewis, Patrick SH and Wu, Ledell and Edunov, Sergey and Chen, Danqi and Yih, Wen-tau},
  booktitle={EMNLP'20},
  pages={6769--6781},
  year={2020}
}

@inproceedings{zhang2024iterative,
  title={An Iterative Utility Judgment Framework via LLMs Inspired by Relevance in Philosophy},
  author={Zhang, Hengran and Bi, Keping and Guo, Jiafeng and Cheng, Xueqi},
  booktitle={Findings of the ACL 2026},
  year={2026}
}

@article{robertson1977probability,
  title={The probability ranking principle in IR},
  author={Robertson, Stephen E},
  journal={Journal of documentation},
  volume={33},
  number={4},
  pages={294--304},
  year={1977},
  publisher={MCB UP Ltd}
}

@inproceedings{zhu2012more,
  title={More than relevance: high utility query recommendation by mining users' search behaviors},
  author={Zhu, Xiaofei and Guo, Jiafeng and Cheng, Xueqi and Lan, Yanyan},
  booktitle={CIKM'12},
  pages={1814--1818},
  year={2012}
}

@inproceedings{lu2018between,
  title={Between clicks and satisfaction: Study on multi-phase user preferences and satisfaction for online news reading},
  author={Lu, Hongyu and Zhang, Min and Ma, Shaoping},
  booktitle={SIGIR'18},
  pages={435--444},
  year={2018}
}

@inproceedings{luo2017does,
  title={Does document relevance affect the searcher's perception of time?},
  author={Luo, Cheng and Liu, Yiqun and Sakai, Tetsuya and Zhou, Ke and Zhang, Fan and Li, Xue and Ma, Shaoping},
  booktitle={WSDM'17},
  pages={141--150},
  year={2017}
}

@inproceedings{dai2020u,
  title={U-rank: Utility-oriented learning to rank with implicit feedback},
  author={Dai, Xinyi and Hou, Jiawei and Liu, Qing and Xi, Yunjia and Tang, Ruiming and Zhang, Weinan and He, Xiuqiang and Wang, Jun and Yu, Yong},
  booktitle={CIKM'20},
  pages={2373--2380},
  year={2020}
}

@inproceedings{yang2018hotpotqa,
  title={HotpotQA: A dataset for diverse, explainable multi-hop question answering},
  author={Yang, Zhilin and Qi, Peng and Zhang, Saizheng and Bengio, Yoshua and Cohen, William and Salakhutdinov, Ruslan and Manning, Christopher D},
  booktitle={EMNLP'18},
  pages={2369--2380},
  year={2018}
}

@inproceedings{zhang2023relevance,
  title={From relevance to utility: Evidence retrieval with feedback for fact verification},
  author={Zhang, Hengran and Zhang, Ruqing and Guo, Jiafeng and de Rijke, Maarten and Fan, Yixing and Cheng, Xueqi},
  booktitle={Findings of the EMNLP 2023},
  pages={6373--6384},
  year={2023}
}

@inproceedings{ke2024bridging,
  title={Bridging the preference gap between retrievers and llms},
  author={Ke, Zixuan and Kong, Weize and Li, Cheng and Zhang, Mingyang and Mei, Qiaozhu and Bendersky, Michael},
  booktitle={ACL'24},
  pages={10438--10451},
  year={2024}
}

@article{gan2024similarity,
  title={Similarity is not all you need: Endowing retrieval augmented generation with multi layered thoughts},
  author={Gan, Chunjing and Yang, Dan and Hu, Binbin and Zhang, Hanxiao and Li, Siyuan and Liu, Ziqi and Shen, Yue and Ju, Lin and Zhang, Zhiqiang and Gu, Jinjie and others},
  journal={arXiv preprint arXiv:2405.19893},
  year={2024}
}

@article{hofstatter2022multi,
  title={Multi-Task Retrieval-Augmented Text Generation with Relevance Sampling},
  author={Hofst{\"a}tter, Sebastian and Chen, Jiecao and Raman, Karthik and Zamani, Hamed},
  journal={arXiv preprint arXiv:2207.03030},
  year={2022}
}

@article{dai2025seper,
  title={Seper: Measure retrieval utility through the lens of semantic perplexity reduction},
  author={Dai, Lu and Xu, Yijie and Ye, Jinhui and Liu, Hao and Xiong, Hui},
  journal={ICLR'26},
  year={2025}
}

@article{zhao2024seer,
  title={Seer: Self-aligned evidence extraction for retrieval-augmented generation},
  author={Zhao, Xinping and Li, Dongfang and Zhong, Yan and Hu, Boren and Chen, Yibin and Hu, Baotian and Zhang, Min},
  journal={EMNLP'24},
  year={2024}
}

@inproceedings{gao2024preference,
  title={Preference-Guided Refactored Tuning for Retrieval Augmented Code Generation},
  author={Gao, Xinyu and Xiong, Yun and Wang, Deze and Guan, Zhenhan and Shi, Zejian and Wang, Haofen and Li, Shanshan},
  booktitle={ASE'24},
  pages={65--77},
  year={2024}
}

@inproceedings{hu2023read,
  title={Read it twice: Towards faithfully interpretable fact verification by revisiting evidence},
  author={Hu, Xuming and Hong, Zhaochen and Guo, Zhijiang and Wen, Lijie and Yu, Philip},
  booktitle={SIGIR'23},
  pages={2319--2323},
  year={2023}
}

@article{izacard2020distilling,
  title={Distilling knowledge from reader to retriever for question answering},
  author={Izacard, Gautier and Grave, Edouard},
  journal={ICLR'21},
  year={2020}
}

@article{izacard2022few,
author = {Izacard, Gautier and Lewis, Patrick and Lomeli, Maria and Hosseini, Lucas and Petroni, Fabio and Schick, Timo and Dwivedi-Yu, Jane and Joulin, Armand and Riedel, Sebastian and Grave, Edouard},
title = {Atlas: few-shot learning with retrieval augmented language models},
year = {2023},
issue_date = {January 2023},
publisher = {JMLR.org},
volume = {24},
number = {1},
issn = {1532-4435},
journal = {J. Mach. Learn. Res.},
month = jan,
articleno = {251},
numpages = {43},
}

@article{lewis2020retrieval,
  title={Retrieval-augmented generation for knowledge-intensive nlp tasks},
  author={Lewis, Patrick and Perez, Ethan and Piktus, Aleksandra and Petroni, Fabio and Karpukhin, Vladimir and Goyal, Naman and K{\"u}ttler, Heinrich and Lewis, Mike and Yih, Wen-tau and Rockt{\"a}schel, Tim and others},
  journal={NeurIPS'20},
  volume={33},
  pages={9459--9474},
  year={2020}
}

@inproceedings{su2403dragin,
  title={Dragin: Dynamic retrieval augmented generation based on the real-time information needs of large language models},
  author={Su, Weihang and Tang, Yichen and Ai, Qingyao and Wu, Zhijing and Liu, Yiqun},
  booktitle={ACL'24},
  pages={12991--13013},
  year={2024}
}

@article{zheng2025deepresearcher,
  title={Deepresearcher: Scaling deep research via reinforcement learning in real-world environments},
  author={Zheng, Yuxiang and Fu, Dayuan and Hu, Xiangkun and Cai, Xiaojie and Ye, Lyumanshan and Lu, Pengrui and Liu, Pengfei},
  journal={arXiv preprint arXiv:2504.03160},
  year={2025}
}

@article{zhang2025evolvesearch,
  title={EvolveSearch: An Iterative Self-Evolving Search Agent},
  author={Zhang, Dingchu and Zhao, Yida and Wu, Jialong and Li, Baixuan and Yin, Wenbiao and Zhang, Liwen and Jiang, Yong and Li, Yufeng and Tu, Kewei and Xie, Pengjun and others},
  journal={arXiv preprint arXiv:2505.22501},
  year={2025}
}

@article{zhang2025unleashing,
  title={Unleashing the Power of LLMs in Dense Retrieval with Query Likelihood Modeling},
  author={Zhang, Hengran and Bi, Keping and Guo, Jiafeng and Sun, Xiaojie and Liu, Shihao and Shi, Daiting and Yin, Dawei and Cheng, Xueqi},
  journal={arXiv preprint arXiv:2504.05216},
  year={2025}
}

@inproceedings{jiang2023active,
  title={Active retrieval augmented generation},
  author={Jiang, Zhengbao and Xu, Frank F and Gao, Luyu and Sun, Zhiqing and Liu, Qian and Dwivedi-Yu, Jane and Yang, Yiming and Callan, Jamie and Neubig, Graham},
  booktitle={EMNLP'23},
  pages={7969--7992},
  year={2023}
}

@inproceedings{shao2023enhancing,
    title = "Enhancing Retrieval-Augmented Large Language Models with Iterative Retrieval-Generation Synergy",
    author = "Shao, Zhihong  and
      Gong, Yeyun  and
      Shen, Yelong  and
      Huang, Minlie  and
      Duan, Nan  and
      Chen, Weizhu",
    booktitle = "Findings of the EMNLP 2023",
    month = dec,
    year = "2023",
    pages = "9248--9274",
}

@article{li2023llatrieval,
  title={Llatrieval: Llm-verified retrieval for verifiable generation},
  author={Li, Xiaonan and Zhu, Changtai and Li, Linyang and Yin, Zhangyue and Sun, Tianxiang and Qiu, Xipeng},
  journal={arXiv preprint arXiv:2311.07838},
  year={2023}
}

@inproceedings{jeong2024adaptive,
    title = "Adaptive-{RAG}: Learning to Adapt Retrieval-Augmented Large Language Models through Question Complexity",
    author = "Jeong, Soyeong  and
      Baek, Jinheon  and
      Cho, Sukmin  and
      Hwang, Sung Ju  and
      Park, Jong",
    booktitle = "NAACL'24",
    month = jun,
    year = "2024",
    address = "Mexico City, Mexico",
    publisher = "Association for Computational Linguistics",
    pages = "7036--7050",
}

@article{gao2024smartrag,
  title={Smartrag: Jointly learn rag-related tasks from the environment feedback},
  author={Gao, Jingsheng and Li, Linxu and Li, Weiyuan and Fu, Yuzhuo and Dai, Bin},
  journal={ICLR'25},
  year={2025}
}

@inproceedings{zhang2025distilling,
  title={Distilling a Small Utility-Based Passage Selector to Enhance Retrieval-Augmented Generation},
  author={Zhang, Hengran and Bi, Keping and Guo, Jiafeng and Zhang, Jiaming and Wang, Shuaiqiang and Yin, Dawei and Cheng, Xueqi},
  booktitle={SIGIR-AP'25},
  pages={22--30},
  year={2025}
}

@article{zhang2025llm,
  title={LLM-Specific Utility: A New Perspective for Retrieval-Augmented Generation},
  author={Zhang, Hengran and Bi, Keping and Guo, Jiafeng and Zhang, Jiaming and Wang, Shuaiqiang and Yin, Dawei and Cheng, Xueqi},
  journal={arXiv preprint arXiv:2510.11358},
  year={2025}
}

@inproceedings{zhang2025leveraging,
  title={Utility-Focused LLM Annotation for Retrieval and Retrieval-Augmented Generation},
  author={Zhang, Hengran and Tang, Minghao and Bi, Keping and Guo, Jiafeng and Liu, Shihao and Shi, Daiting and Yin, Dawei and Cheng, Xueqi},
  booktitle={EMNLP'25},
  pages={1683--1702},
  year={2025}
}

@inproceedings{jain2025modeling,
  title={Modeling Contextual Passage Utility for Multihop Question Answering},
  author={Jain, Akriti and Garimella, Aparna},
  booktitle={IJCNLP-AACL'25},
  pages={464--471},
  year={2025}
}

@article{jin2025search,
  title={Search-r1: Training llms to reason and leverage search engines with reinforcement learning},
  author={Jin, Bowen and Zeng, Hansi and Yue, Zhenrui and Yoon, Jinsung and Arik, Sercan and Wang, Dong and Zamani, Hamed and Han, Jiawei},
  journal={arXiv preprint arXiv:2503.09516},
  year={2025}
}

@inproceedings{wang2016beyond,
  title={Beyond ranking: Optimizing whole-page presentation},
  author={Wang, Yue and Yin, Dawei and Jie, Luo and Wang, Pengyuan and Yamada, Makoto and Chang, Yi and Mei, Qiaozhu},
  booktitle={WSDM2016},
  pages={103--112},
  year={2016}
}

@inproceedings{yu2016dynamic,
  title={A dynamic recurrent model for next basket recommendation},
  author={Yu, Feng and Liu, Qiang and Wu, Shu and Wang, Liang and Tan, Tieniu},
  booktitle={SIGIR'16},
  pages={729--732},
  year={2016}
}

@inproceedings{canale-etal-2025-bes4rag,
    title = "{BES}4{RAG}: A Framework for Embedding Model Selection in Retrieval-Augmented Generation",
    author = "Canale, Lorenzo  and
      Scotta, Stefano  and
      Messina, Alberto  and
      Farinetti, Laura",
    booktitle = "CLiC-it 2025",
    month = sep,
    year = "2025",
    address = "Cagliari, Italy",
    publisher = "CEUR Workshop",
    pages = "134--142",
    ISBN = "979-12-243-0587-3"
}
\end{document}